\documentclass[journal]{vgtc}                     % final (journal style)
\onlineid{1427}

\vgtccategory{Research}

\vgtcpapertype{algorithm}

\newcommand{\sysname}{\textit{\textbf{P}arams\textbf{D}rag}}
\newcommand{\clb}{}

\title{\sysname{}: Interactive Parameter Space Exploration via Image-Space Dragging}

\author{%
  \authororcid{Guan Li $^{1,2}$}{0000-0001-6436-3650},
  \authororcid{Yang Liu $^{3}$}{0009-0003-9893-8696},
  \authororcid{Guihua Shan* $^{1,2,4}$}{0000-0002-8283-2278},
  \authororcid{Shiyu Cheng $^{1}$}{0000-0002-7400-515X},
  \authororcid{Weiqun Cao $^{3}$}{0000-0001-6195-6928},
  \\
  \authororcid{Junpeng Wang $^{5}$}{0000-0002-1130-9914},
  and 
  \authororcid{Ko-Chih Wang $^{6}$}{0000-0002-7241-1939}
}

\authorfooter{
  \item $^1$Computer Network Information Center, Chinese Academy of Sciences
  \item $^2$University of Chinese Academy of Sciences
  \item $^3$Beijing Forestry University
  \item $^4$Hangzhou Institute for Advanced Study, UCAS
  \item $^5$Visa Research
  \item $^6$National Taiwan Normal University
  \item Guihua Shan is the corresponding author. E-mail: sgh@sccas.cn
}

\abstract{Numerical simulation serves as a cornerstone in scientific modeling, yet the process of fine-tuning simulation parameters poses significant challenges. 
Conventionally, parameter adjustment relies on extensive numerical simulations, data analysis, and expert insights, resulting in substantial computational costs and low efficiency. 
The emergence of deep learning in recent years has provided promising avenues for more efficient exploration of parameter spaces.
However, existing approaches often lack intuitive methods for precise parameter adjustment and optimization.
To tackle these challenges, we introduce \sysname{}, a model that facilitates parameter space exploration through direct interaction with visualizations. Inspired by DragGAN, our \sysname{} model operates in three steps. First, the generative component of \sysname{} generates visualizations based on the input simulation parameters. Second, by directly dragging structure-related features in the visualizations, users can intuitively understand the controlling effect of different parameters. Third, with the understanding from the earlier step, users can steer \sysname{} to produce dynamic visual outcomes. Through experiments conducted on real-world simulations and comparisons with state-of-the-art deep learning-based approaches, we demonstrate the efficacy of our solution.
}


\graphicspath{{figs/}{figures/}{pictures/}{images/}{./}} % where to search for the images

\usepackage{tabu}                      % only used for the table example
\usepackage{booktabs}                  % only used for the table example
\usepackage{lipsum}                    % used to generate placeholder text
\usepackage{mwe}                       % used to generate placeholder figures

\usepackage{mathptmx}                  % use matching math font
\usepackage{amsmath}
\usepackage{algorithm}
\usepackage{algorithmic}
\usepackage{multirow}
\usepackage{makecell}
\usepackage{gensymb}

\begin{document}

%%%%%%%%%%%%%%%%%%%%%%%%%%%%%%%%%%%%%%%%%%%%%%%%%%%%%%%%%%%%%%%%
%%%%%%%%%%%%%%%%%%%%%% START OF THE PAPER %%%%%%%%%%%%%%%%%%%%%%
%%%%%%%%%%%%%%%%%%%%%%%%%%%%%%%%%%%%%%%%%%%%%%%%%%%%%%%%%%%%%%%%

\firstsection{Introduction}
\maketitle
Numerical simulations are integral to modern scientific inquiry, bridging theoretical models and real-world phenomena. These simulations enable scientists to investigate complex physical processes, verify scientific hypotheses, and refine the underlying physical models. In scientific research, the iterative process of conducting multiple simulation runs with varying simulation parameter conditions is essential for exploring the uncertainties inherent in physical parameters. For instance, in cosmological simulations, by performing a multitude of simulations and analyzing under a wide range of initial conditions, scientists can gain insights into the formation and evolution of the universe and the impact of different initial conditions on structural formation.

However, optimizing simulation parameters is a complex and computationally resource-intensive task. Each parameter adjustment requires rerunning the simulation and evaluating its effects, a process that demands significant computational power and substantial storage for the resulting data. Furthermore, exploring valuable insights from these extensive datasets necessitates visualization and in-depth analysis of each simulation run's outcomes. Scientists refine simulation parameters by analyzing the key features and structures revealed in these results, moving closer to achieving their research goals.
For example, scientists often employ simulations to repeat the formation of galaxy clusters or dark matter halos in studying the universe's large-scale structures. They may need to repeatedly adjust initial conditions and other relevant parameters to produce structures and physical properties corresponding to observational data. This process includes not only fine-tuning the parameters but also continuously monitoring the simulation outcomes. Scientists usually require this process to adjust simulation parameters to facilitate the formation of the anticipated structures.

The advent of deep learning models has introduced novel approaches to parameter exploration in numerical simulations. Recent research has introduced surrogate models, such as InSituNet~\cite{he2019insitunet}, GNN-Surrogate~\cite{shi2022gnn}, and VDL-Surrogate~\cite{shi2022vdl}, leveraging deep neural networks to boost the efficiency of this exploration. 
These models employ deep learning to establish a mapping from initial simulation parameters to the ultimate outcomes, enabling direct result prediction and circumventing the need for time-consuming numerical simulations. 
However, these methods have yet to address a critical issue: \textbf{intuitively interacting with predicted outcomes and retrieving the input simulation parameters corresponding to the outcomes.} 
Existing surrogate models primarily aim to predict outcomes based on parameter adjustments. If the predicted results fall short of expectations, experts must infer the direction of parameter adjustments by drawing on their expertise. This situation leaves the parameter adjustment process needing a more precise and controllable method, thereby suggesting that the efficiency of exploring the parameter space needs enhancement.

To support the intuitive simulation parameter exploration, we propose \sysname{}, a model that facilitates parameter space exploration through direct interaction with visualizations. Our work has two main goals: (1) to enable scientists to drag a feature of interest to the desired location on the visualization image and generate the corresponding image; (2) to retrieve the corresponding simulation parameter conditions of the generated images. 
Inspired by DragGAN~\cite{pan2023drag}, a technique to support interactive manipulation on classic image generation applications, we propose an approach that aims to properly edit the latent vector of our deep-learning-based surrogate model to support scientists dragging a feature on the visualization and then generate the corresponding image.
However, a fundamental difference between applications of classic image generation and scientific data visualization generation leads to a few technical challenges. The difference is that the scientific visualization generation focuses on the accuracy of the generated visualization corresponding to the provided simulation parameter condition. Conversely, classic image generation applications focus on producing natural and sharp images, making it challenging to abstract comprehensive and quantifiable labels.
Our study shows that the valid latent vectors of a well-trained scientific visualization generation model are distributed more discretely than classic image generation models. This phenomenon prohibits directly editing latent vectors as the classic image generation model in the latent space. 
In our work, we develop a deep-learning-based surrogate model that takes simulation parameters as input and generates the corresponding scientific visualization. Furthermore, we propose a gradient descent-based parameter search algorithm for editing the latent vector, enabling stepwise movement only on valid points in the discrete latent space. This algorithm ensures that only valid images are generated and also guarantees smooth transitions between images as scientists manipulate features on the image. Lastly, a function is provided to retrieve the corresponding simulation parameters of any generated image as scientists adjust features on the visualization.
Overall, this work makes the following four contributions:

\begin{itemize}
\item We elucidate the differences in latent vector distributions and the variations in editing methodologies between classic image generation and scientific visualization image generation. 

\item We design a deep-learning-based surrogate model that inputs simulation parameters to generate corresponding scientific data visualizations efficiently.

\item We define the structure-based patch and implement direct interaction with visualization through feature supervision and tracking.

\item We implement a gradient descent-based parameter search method to support the generation of dynamic images and the inversion of corresponding parameters for visualization images.

\end{itemize}

\section{Related Work}
This section reviews the relevant work on applying deep learning in scientific visualization and analyzing the parameter space in numerical simulations.

\subsection{Deep Learning for Scientific Visualization}
Deep learning techniques are increasingly demonstrating their significant potential in scientific visualization. Much research has utilized deep learning to address challenges within scientific visualization, achieving good results. The tasks related to our study primarily fall into prediction and visualization generation~\cite{wang2022dl4scivis}.

Prediction primarily involves using deep learning models to learn from existing data distributions and forecast new data instances. For instance, in data prediction, CECAV-DNN~\cite{he2020cecav} employed a DNN model to predict ensemble similarity, facilitating the analysis of ensemble data. 
Tkachev \textit{et al.}~\cite{tkachev2019local} utilized neural networks to predict voxel values and identify anomalous data. 
Both Hong \textit{et al.}~\cite{hong2018access} and Han \textit{et al.}~\cite{han2022exploratory} applied deep learning models for predicting particle-related data. 
Regarding visualization prediction, the works of Yang \textit{et al.}~\cite{yang2019deep} and Shi \textit{et al.}~\cite{shi2019cnns} leveraged neural networks to predict visualization images under new parameters, supporting the evaluation of viewpoint parameters. 
Engel and Ropinski~\cite{engel2020deep} introduced a deep-learning approach for predicting per-voxel ambient occlusion in volumetric datasets within direct volume rendering.

The task of visualization generation involves using neural networks to learn the mapping from input parameters to visualization outputs, thereby enabling the generation of visualization results under new parameters. Its objective is to replace traditional, potentially cumbersome processes, thus enhancing the efficiency of exploration and analysis. 
This process encompasses different types of input parameters, such as simulation parameters, viewpoint parameters, and visual mapping parameters. 
For example, InSituNet~\cite{he2019insitunet}, GNN-Surrogate~\cite{shi2022gnn}, and VDL-Surrogate~\cite{shi2022vdl} have all replaced simulation programs with deep learning models to improve the efficiency of parameter space exploration, focusing primarily on simulation parameters. 
In viewpoint parameters, DNN-VolVis~\cite{hong2019dnn} achieved rapid rendering under new viewpoint parameters, while Berger \textit{et al.}~\cite{berger2018generative} utilized GANs to acquire visualization images from new viewpoints quickly. 
Regarding visual mapping parameters, DeepDVR~\cite{weiss2021deep} generated similar direct volume rendering results from examples in image space, eliminating the need for explicit feature design and manual transfer function specification. 
Additionally, Weiss \textit{et al.}~\cite{weiss2019volumetric} employed FRVSR-Net to transform low-resolution isosurface rendering images into high-resolution ones, and in~\cite{weiss2020learning}, they explored the relationship between data and image generation using an end-to-end neural rendering framework.

The first goal of our work is to utilize deep neural networks to predict the visualization results from parameters. These related studies provide a solid theoretical foundation for our work, offering valuable insights for model design and loss function construction.

\subsection{Parameter Space Analysis}
Visualization techniques are increasingly playing a pivotal role in the analysis of Parameter Spaces~\cite{wang2018visualization}. Existing related work can be technically categorized into two types: the first is traditional parameter analysis methods, and the second is surrogate model approaches~\cite{shi2022vdl}.

In traditional analysis methods, the process typically starts with running simulation programs multiple times and then collecting the simulation data for subsequent visual analysis. In this paradigm, visualization focuses on exploring better design or data processing algorithms to support uncertainty visualization. 
For instance, employing curve-related visual forms to analyze spatiotemporal uncertainty in ensemble data~\cite{ferstl2016visual, ferstl2015streamline, mirzargar2014curve, whitaker2013contour}, modeling data uncertainty using probability density functions for further visual analysis~\cite{thompson2011analysis, bensema2015modality, liu2012gaussian, otto2012vortex, pfaffelmoser2012visualization, pothkow2011probabilistic, hazarika2017uncertainty}, and some studies explore visualizing high-dimensional data to support uncertainty analysis, such as scatter plots~\cite{matkovic2009interactive, orban2018drag, splechtna2015interactive}, parallel coordinate plots~\cite{obermaier2015visual, wang2016multi}, and matrix charts~\cite{poco2014visual, sanyal2010noodles, jarema2015comparative}. 
However, a significant drawback of traditional methods was their substantial computational resource requirements to support the running of simulation programs, leading to inefficient parameter analysis and optimization.

Machine learning methods provided new approaches to exploring parameter spaces, with the use of machine learning models to replace simulation programs significantly enhancing the efficiency of parameter exploration.
For instance, NNVA~\cite{hazarika2019nnva} utilized a neural network-based surrogate model to facilitate the interactive exploration of high-dimensional input parameter spaces in yeast cell polarization simulations.  
Alden \textit{et al.}~\cite{alden2018using} employed surrogate models to efficiently analyze biological systems simulations, enabling advanced statistical analyses and optimizations that were previously infeasible due to computational limitations.
Similarly, Erdal \textit{et al.}~\cite{erdal2020sampling} applied Gaussian Process Emulators to enhance the efficiency of sampling in ensemble-based sensitivity analysis of environmental models, demonstrating a substantial improvement in identifying behavioral samples and local sensitivities.
Furthermore, research closely aligned with our work includes InSituNet~\cite{he2019insitunet}, GNN-Surrogate~\cite{shi2022gnn}, and VDL-Surrogate~\cite{shi2022vdl}, all of which utilize neural networks to substitute for numerical simulation programs. 
These related studies focused on adjusting parameters to modify the final outcomes, yet they often required multiple modifications and iterative analyses when the predicted results failed to meet expectations. This process still largely depended on expert intuition for parameter adjustments, indicating that there is still room for improvement in the efficiency of exploring the parameter space.

\section{Background}
Generative models constitute a distinct class within the area of deep learning, fundamentally focused on learning data distributions. These models are capable of generating new data instances by learning the data distribution. Image generation is one of the tasks for generative models, which aims to create high-quality and lifelike images. Among the models frequently used for this purpose are Autoencoders (AE)~\cite{hinton2006reducing}, Variational Autoencoders (VAE)~\cite{kingma2013auto}, and Generative Adversarial Networks (GANs)~\cite{goodfellow2014generative}. 
The task of generating scientific visualization images differs from classic generative tasks.

In computer vision, classic generative tasks primarily aim to produce images from the real world. However, it is difficult to assign descriptions with precise labels to these images, leading to the reliance on ambiguous labels or introducing noise as input. 
For instance, within GANs, random noise is employed as the initial input to the generator. This noise undergoes several transformations across the network's layers to yield samples that mimic the appearance of images from the actual dataset. Furthermore, in the case of StyleGAN~\cite{karras2019style} and StyleGAN2~\cite{karras2020analyzing}, noise is not merely an input for the generator but also plays a crucial role in refining the latent space vectors. This application of noise serves to inject random details into the images, thus enhancing the model's ability to generate diverse data. 
Although incorporating noise in the latent space can enrich the model's generative, an input label can potentially map and generate multiple distinct images.

In the field of scientific visualization, the generation of images under the constraints of simulation parameters presents unique challenges. The visualization image of scientific data is tied to parameters with distinct meanings, such as physical or visualization parameters. This connection means that every visualization is associated with a unique and explicit label, such as simulation parameter values. 
As a result, generative models designed within these parameter constraints strive to remove ambiguity from the input to the output process, typically preventing from introduction of noise into the model. This scheme has been applied to existing deep-learning scientific visualization generators like InSituNet, GNN-Surrogate, and VDL-Surrogate. 
These models learn the mapping between simulation parameters and visualization outcomes to support parameter space exploration.

Generative models exhibit significant differences when employing explicit labels (such as simulation parameters) versus ambiguous labels (such as category labels mixed noise). 
This distinction affects not only the models' structures but also the distributions within their latent spaces. Consequently, this variation often makes traditional latent space editing techniques unsuitable for use in scenarios with explicit labeling.
This study aims to develop generative models constrained by parameters, utilizing these models to enhance user interaction and analyze predictive outcomes.
In the following sections, we introduce the proposed generative models under simulation parameter constraints and the methods for editing the corresponding latent vectors.

\section{Overview}
\begin{figure}[tb]
  \centering 
  \includegraphics[width=\columnwidth]{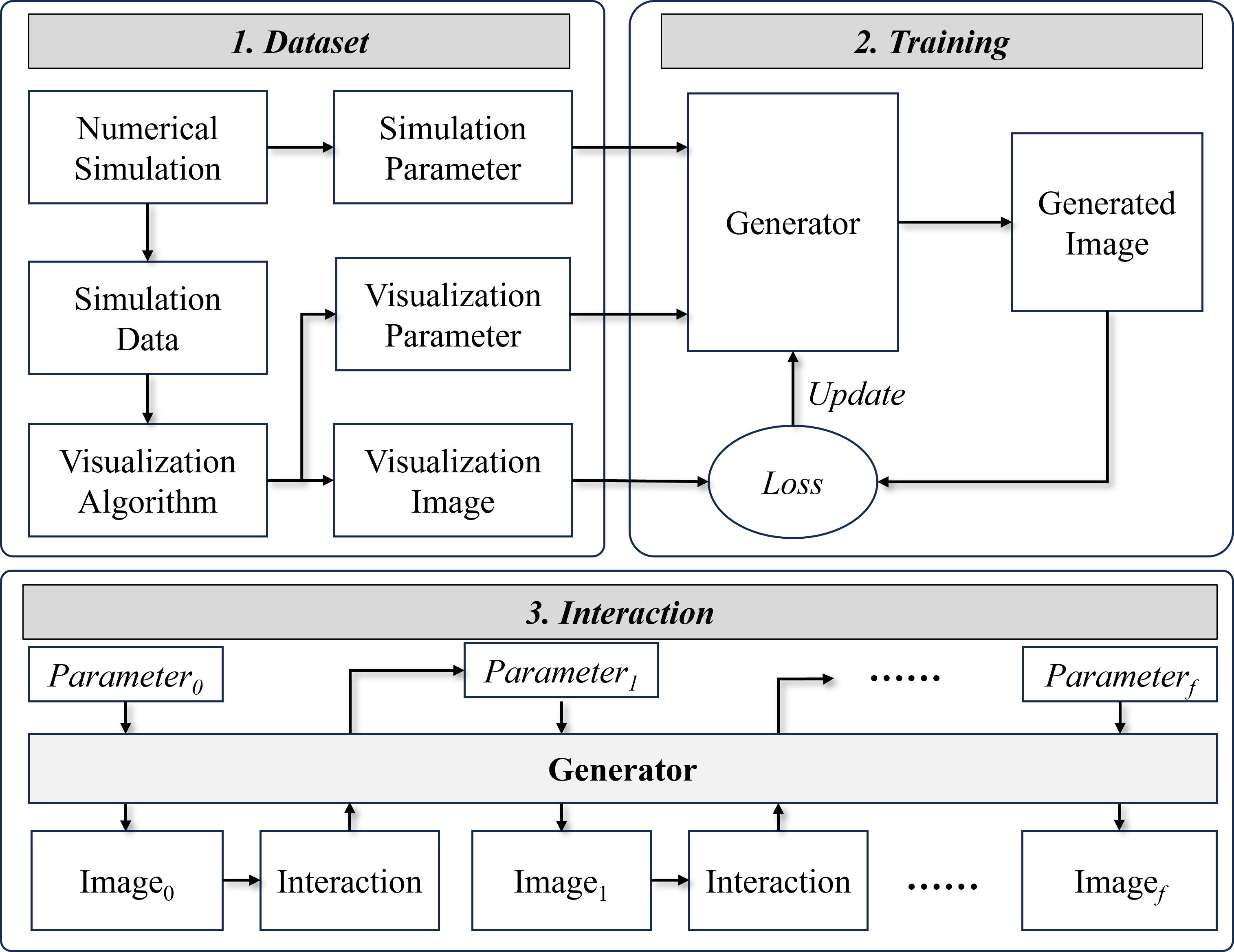}
  \caption{The high-level pipeline of \sysname{}.
  }
  \label{fig:pipeline}
\end{figure}

Our objective is to enable experts to interact with predicted images and get the corresponding simulation parameters, thus improving the efficiency of optimizing simulation parameters. 
Therefore, a generator is required to predict visualization outcomes by input simulation parameters. 
Based on the generator, we also propose an algorithm that enables users to intuitively drag the feature of interest on the image to generate the corresponding visualization and derive simulation parameters.

Figure~\ref{fig:pipeline} illustrates the workflow of \sysname{}, which integrates three main components: the construction of the dataset, the training of the generator, and interaction with the visualization image.
During the phase of constructing the dataset, we visualize simulation data and store both the simulation parameters and visualization parameters associated with each image to build the dataset. 
In the subsequent phase, we refine the generator's architecture through a state-of-the-art image generation model to fit our scientific visualization purpose. The generator is trained to learn the mapping between the simulation parameters and visualization images using the above dataset. The trained generator can predict the image of given simulation parameters as input.
The final phase is interactive image prediction and simulation parameter derivation through user interactions. We define the concept of the structure-based patch, which enables users to select features of interest within the visualizations. Upon interaction, feature supervision and feature tracking guide the generator in producing an image sequence, which illustrates the transition before and after user interactions. A gradient-based search technique is also proposed to derive the parameters associated with each image. 
By integrating these features, \sysname{} model implements a novel approach to parameter space exploration that enables direct interaction with the predicted visualization images.

Additionally, we also demonstrate the differences in the generative models' latent vector distributions under applications of classic image generation and scientific visualization generation. We show how these differences impact the strategies for editing latent space vectors. 
This exploration reveals the rationale behind our model's structural design, showcasing how distinct approaches to handling latent space can impact model performance and editing capabilities, particularly in scenarios with scientific simulation parameter space exploration. 

\section{\sysname{} Model}
This section introduces the technical details of \sysname{} model.

\subsection{Dataset Construction}
The process of constructing the training set commences with the identification of simulation parameters based on expert analysis objectives. After this, these parameters are sampled, followed by the execution of numerical simulations for each parameter set, with the data being recorded. The final step involves employing visualization algorithms on the simulation data to convert this data into images. The data to be stored throughout this procedure includes:

\textbf{Simulation parameters}, formatted as one-dimensional vectors, encompassing physical parameters, time, and other variables pertinent to the simulation program.

\textbf{Visualization parameters}, also in one-dimensional vectors, include aspects such as viewpoint, threshold, and other parameters relevant to the visualization algorithm.

\textbf{Visualization images}, formatted as portable network graphics (PNG), which are generated from the simulation data through visualization algorithms and correspond to the simulation and visualization parameters.

\subsection{Generator}
In this subsection, we introduce our generator's architecture and loss function, which is designed to support latent vector editing and accurate visualization image prediction based on parameters.

\subsubsection{Generator Architecture}
\begin{figure}[tb]
  \centering 
  \includegraphics[width=\columnwidth]{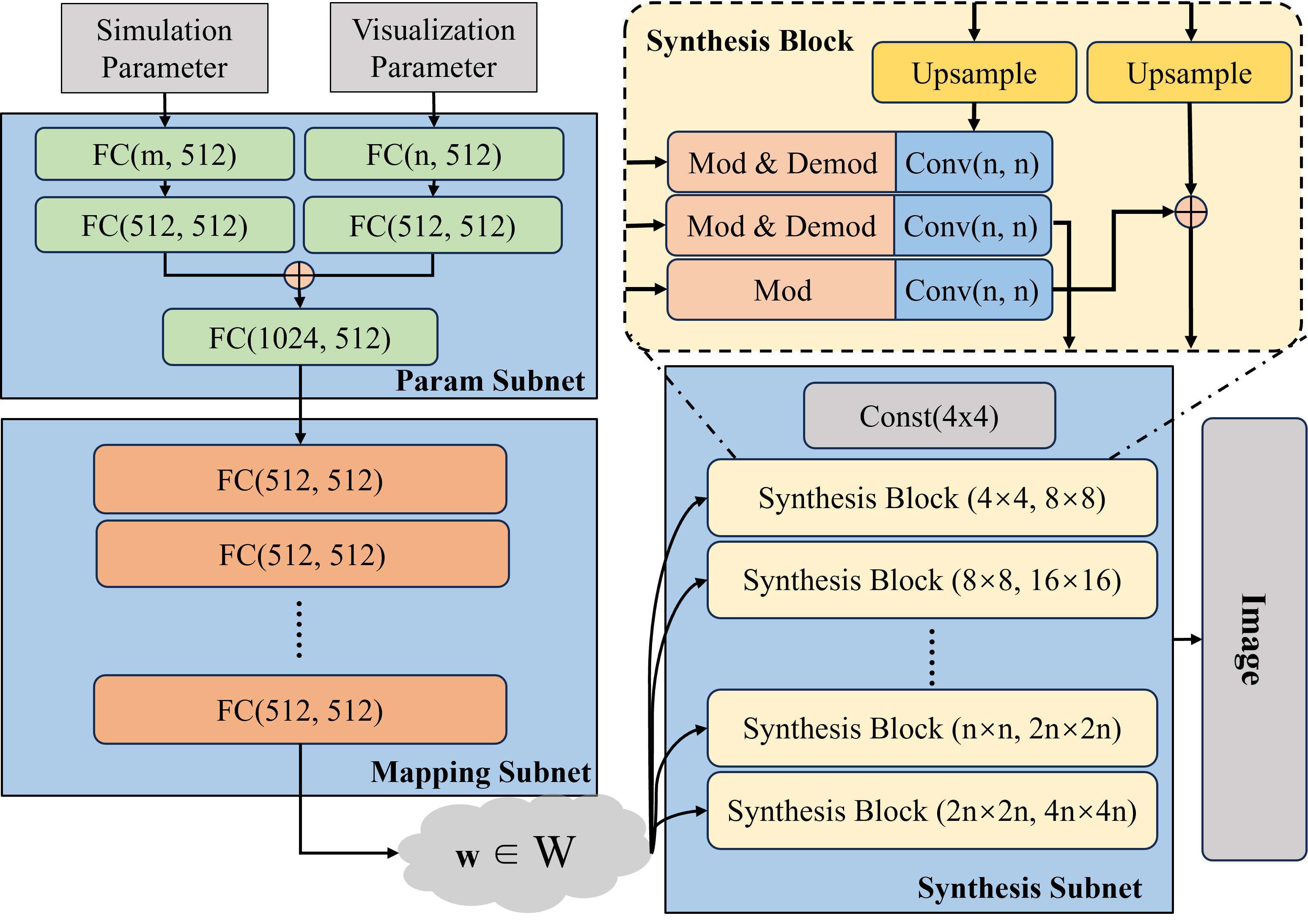}
  \caption{The model structure of the generator, with inputs being the simulation parameters and visualization parameters, and the output being the visualization image. 
  }
  \label{fig:architecture}
\end{figure}

Figure~\ref{fig:architecture} displays the architecture of the generator, where the input consists of simulation parameters and visualization parameters, and the output is a visualization image. 
It comprises three sub-networks: Param Subnet, Mapping Subnet, and Synthesis Subnet. 
We reference multiple state-of-the-art models, such as StyleGAN2 and InSituNet, to guide our network design in the model's design process.

The Param Subnet primarily focuses on extracting features from the input parameters and transforming them into a format usable for the generation process. It comprises two branches: one for handling simulation parameters and the other for dealing with visualization parameters. Each branch features two fully connected layers (FC), designated as $FC(m, 512)$ and $FC(n, 512)$, where $m$ and $n$ represent the dimensions of different types of parameters. We use ReLU~\cite{nair2010rectified} as the activation function in all FC layers. The outputs of these two branches are then merged into a single vector through a concatenation operation, which is further processed by another fully connected layer, $FC(1024, 512)$. The output of this subnet is subsequently provided to the Mapping Subnet.

The Mapping Subnet serves to transform the output of the Param Subnet into a latent space $w$, which often possesses better attributes than the original parameter space, such as being easier for subsequent networks to learn and disentangle. 
The Mapping Subnet typically consists of multiple fully connected layers, with each layer in our model being $FC(512, 512)$, and multiple such layers can be stacked together. 
The purpose of each layer is to transform the input features, enhancing their distribution characteristics gradually. 
This sequence of FC and ReLU layers enables the network to learn complex nonlinear mappings, ultimately producing the latent vector $w$.

The Synthesis Subnet generates the final image from the latent vector $w$, which is output by the Mapping Subnet. In developing the Synthesis Subnet, we drew inspiration from the architectural framework of StyleGAN2 and made custom modifications, notably omitting noise injection components. This design change is necessary to ensure that one given parameter input does not produce two distinct visualizations. The details are discussed in Section \ref{sec:theoreticalFoudation}.
The Synthesis Subnet commences with a constant tensor, Const(4x4), as the initial feature map, which then undergoes progressive transformations via multiple Synthesis blocks to incrementally enhance the image resolution. 
Within each Synthesis block, an upsampling procedure doubles the feature map's dimensions, followed by convolutional operations refining the features. 
Moreover, the Synthesis blocks incorporate Modulation (Mod) and Demodulation (Demod) operations. 
Modulation dynamically adjusts the convolutional kernels using the latent vector $w$, enabling the network to control the representation of image features at various levels.
Demodulation normalizes the modulated weights by computing a standardization coefficient, thus mitigating the risk of excessive activation values resulting from the modulation process. 
The definitions and implementations of Mod and Demod align with those delineated in StyleGAN2~\cite{karras2020analyzing}. 
These collaborative operations harness the information within $w$ to direct the image synthesis process, culminating in the generating of images with high-quality details.

\subsubsection{Loss Function}
The objective of the loss function constraints is to quantify the differences between the images produced by the generator and the target images, thereby guiding the generator's training.
In this work, we employ three loss functions: content loss ($\mathcal{L}_{content}$), perceptual loss~\cite{johnson2016perceptual} ($\mathcal{L}_{feature}$), and edge loss ($\mathcal{L} _{edge}$), as is shown in Equation~\ref{eq:loss_total}.
\begin{equation}
    \mathcal{L} = \alpha \mathcal{L}_{content}  + \beta \mathcal{L}_{feature} +\gamma \mathcal{L}_{edge}
    \label{eq:loss_total}
\end{equation}
where $\alpha$, $\beta$, and $\gamma$ are the coefficients. 

The content loss measures the pixel-wise difference between the generated and target images. We employ the L1 loss function to fulfill this purpose, as depicted in Equation~\ref{eq:loss_content}. 
\begin{equation}
\mathcal{L}_{content}=\frac{1}{N} {\sum_{i=1}^{N}} \left | {I}_{i} - \hat{I}_{i}\right |
\label{eq:loss_content}
\end{equation}
where $N$ represents the total number of pixels, ${I}_{i}$ is the target value of the $i$th pixel, and $\hat{I}_{i}$ is the predicted value for the $i$th pixel.

The simple difference between pixels does not adequately quantify the differences in image structure features. To capture the distinctions in feature structures within visualized images, we adopt perceptual loss~\cite{johnson2016perceptual}. 
Perceptual loss measures the discrepancy in perceptual quality between two images, thereby making the generated image visually closer to the target image. 
It achieves this by comparing the feature representations at different layers within a pre-trained VGG-19 model~\cite{simonyan2014very}, which can capture information about the content and style of the images, as illustrated in Equation~\ref{eq:loss_feature}.
\begin{equation}
    \mathcal{L}_{\text{feature}} = \sum_{j}^{L} \lambda_{j} \cdot \left \| \phi_{j}(I) - \phi_{j}(\hat{I}) \right \|_{2}^{2}
\label{eq:loss_feature}
\end{equation}
where $\phi_{j}$ represents the activation maps produced by layer $j$, ${I}$ denotes the target images, $\hat{I}$ denotes the predicted images, $\lambda_{j}$ a weighting factor for layer $j$, and $\left \| \cdot \right \|$ is the L2 norm.

In scientific visualization images, the edge features reveal the contours and shapes of the structures. We define an edge loss to ensure the generated images accurately capture these details. The edge loss is based on the Sobel operator for edge extraction, as shown in Equation~\ref{eq:loss_edge}.
\begin{equation}
\mathcal{L}_{edge}=\left \| \psi({I})-\psi(\hat{I}) \right \|_{2}^{2}
\label{eq:loss_edge}
\end{equation}
where $\psi()$ involves applying the Sobel operator through two convolution operations to identify edge regions in an image. 
Edge loss reflects the differences in feature edges within the image, placing special emphasis on the details of lines.

\begin{figure}[tb]
  \centering 
  \includegraphics[width=\columnwidth]{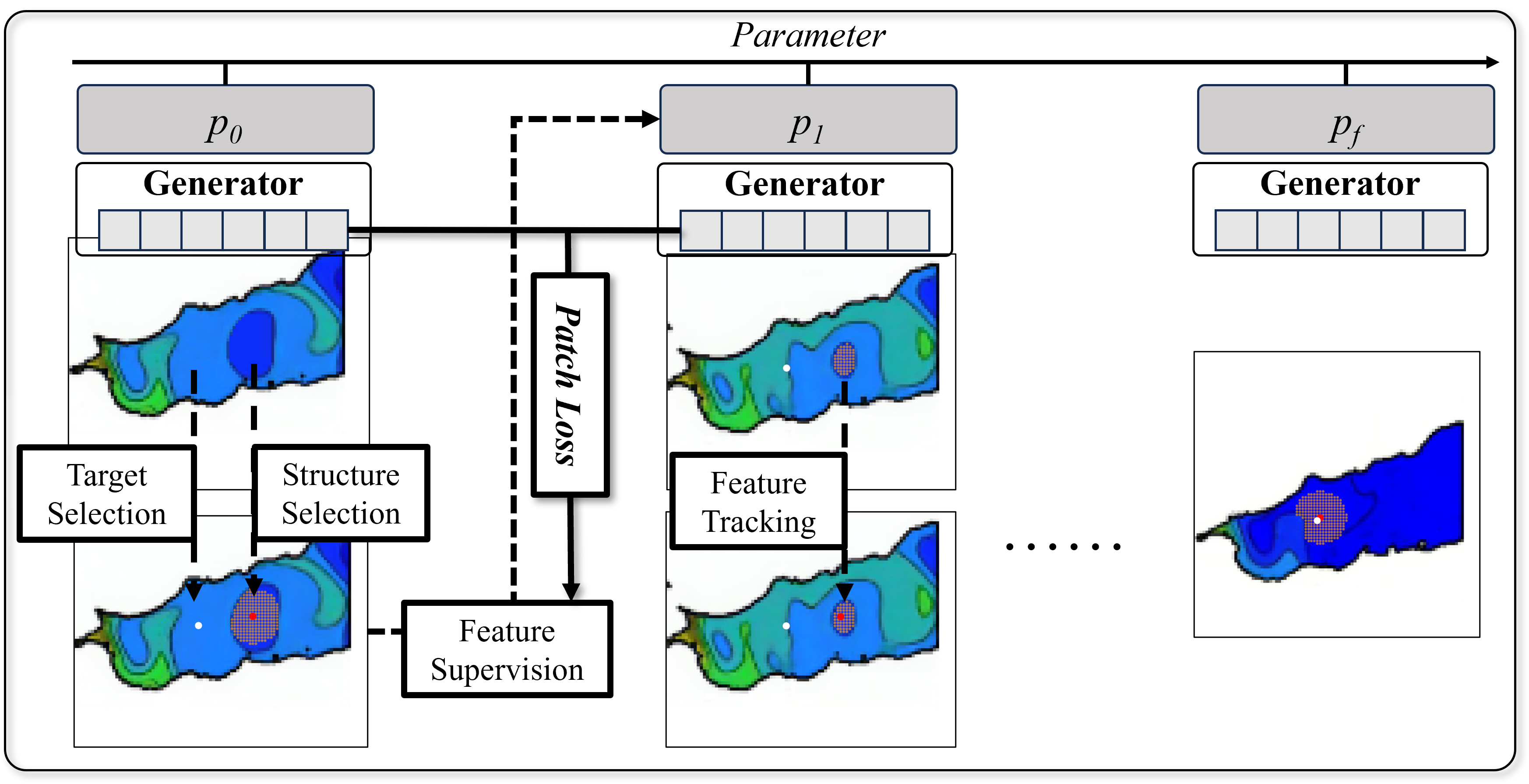}
  \caption{The workflow of interacting with visualization images. Users can directly obtain target parameters and visualization images by selecting feature structures and editing it.}
  \label{fig:drag_pipeline}
\end{figure}

\subsection{Interaction with Visualization Images}
In the interaction part, we draw inspiration from DragGAN~\cite{pan2023drag} to implement the interactive functionality of \sysname{}. Direct application of DragGAN to our generative models for scientific visualization images is impracticable due to the discrete distribution of the model's latent vectors in the latent space, a consequence of explicit label constraints. 
\clb{To address this issue, we improve the method of latent vector editing in DragGAN, enabling our model to perform stepwise changes in the latent space, thereby generating accurate visualization images. The theoretical foundation of our approach is discussed in Section~\ref{sec:theoreticalFoudation}.}

\clb{Figure~\ref{fig:drag_pipeline} illustrates the workflow of our interaction process. Users initially select the interactive objects (i.e., the structure-based patches) and the interaction targets (i.e., intended positions for movement) within the visualization image as input conditions. Subsequently, the loss before and after the structure's movement is computed using a feature supervision method. Following this, the image and parameters are updated utilizing a parameter inversion method based on the loss. Finally, feature tracking is employed to reposition the feature control points. Therefore, a single iteration comprises three critical steps: \textbf{Feature Supervision}, \textbf{Parameter Inversion}, and \textbf{Feature Tracking}. This process repeats until the termination conditions are met. Our method includes two termination conditions: the interactive object reaching the target position and the structural disappearance of the interactive object.}

\subsubsection{Structure-based Patch}
We introduce the concept of the structure-based patch, with the primary goal of assisting users in identifying and selecting specific structures within scientific visualization images.
\clb{Scientific visualization involves converting data attributes into different colors to represent attribute ranges, where multiple pixels of similar colors form a small region. Pixels in this region, having similar attribute values, can be defined as a structure in many domains. For example, a dark halo can be represented as a highlighted local area within the visualization image in cosmological simulations. Therefore, we utilize the structure-based patch to help users locate structures within visualization images and support further editing of these structures.}

The structure-based patch is implemented through a breadth-first search algorithm based on the click location by calculating the pixel similarity $d$ within a radius $r$ of the user's click position. 
\clb{Our method also supports simultaneous selection and editing of multiple feature structures. For any given click position $p_{i}$, the search algorithm is used to construct the pixel set $Patch(p_{i})$. Additionally, a moving target position $g_i$ is set for each click position $p_i$ to guide the direction of the structure's movement. In our implementation, the calculation of structural similarity is conducted under the HSV color model.}

\subsubsection{Feature Supervision}
The objective of feature supervision is to help the movement of selected structures within the visualization image. 
\clb{The essence of feature supervision is to minimize structural changes within the visualization image after movement, ensuring continuous local feature changes in the image.
Therefore, if the feature's corresponding set of pixels $Patch(p_{i})$ is to be moved to a target point $g_i$, this can be achieved by supervising the patch to move a small step $r_m$ towards $g_i$ each time, and then through multiple iterations to accomplish the feature movement.} 
During this process, we employ the loss function for feature supervision from DragGAN, as shown in Equation~\ref{eq:feature_supervison}.
\begin{equation}
\mathcal{L}_{M S} =  \sum_{i=0}^{n} \sum_{q_{i} \in Patch(p_{i})}^{} ||\mathrm{F}(q_{i}) - \mathrm{F}(q_{i} + v_{i})||_{1}
\label{eq:feature_supervison}
\end{equation}
\begin{equation}
v_{i} = \frac{ {g}_{i} - {p}_{i} }{ || {g}_{i} - {p}_{i}  ||_{1}}
\label{eq:direction}
\end{equation}
where n represents the number of selected structures. $q_{i}$ denotes the pixels surrounding the selected structure $Patch(p_{i})$, $v_{i}$ is the vector pointing from the manipulation point to the target point $g_{i}$, $F$ is the feature map from a specific layer chosen within the generator, and $F(q)$ is the feature vector at pixel position $q$.

\subsubsection{Parameter Inversion}
\clb{DragGAN utilizes the first six latent vectors of the $W$ space from StyleGAN2 for gradient search to achieve image editing.
The distribution of latent vectors in the latent space for \sysname{}'s generator differs from that of StyleGAN2's generator, and using a latent vector search approach would result in erroneous images. 
To address this issue, we opt for modifications in the parameter space to ensure the validity of the generated images.
Specifically, we use Equation~\ref{eq:feature_supervison} as the loss function to quantify the differences before and after the patch movement, and then perform a gradient search on the input parameters to achieve editing of the latent vectors.
The theoretical foundation for this approach is discussed in detail in Section~\ref{sec:theoreticalFoudation}.
}

\clb{We implement image editing through gradient search on input parameters, allowing the corresponding parameters of newly generated images to be acquired in real time. Furthermore, based on the diverse requirements of experts, we can conduct gradient searches on multiple input parameters or a single parameter to explore the parameter space under various analytical objectives.}

\subsubsection{Feature Tracking}
The objective of feature tracking is to address the issue of inaccurately tracking control points after the target has moved while also determining whether the feature structure has disappeared. Feature tracking is accomplished through the nearest neighbor search of feature maps, as illustrated in Equation~\ref{eq:feature_tracking}.
\begin{equation}
p_ {i} = \min _{ q_{i}  \in  Square  ( p_{i}, r_{m})} ||\mathrm{F} '( q_{i} )-\mathrm{F} _{0}(p_{i}^{0})||_{1}
\label{eq:feature_tracking}
\end{equation}
where $\mathrm{F} _{0}()$ represents the initial feature map, $p_{i}^{0}$ denotes the initial position selected by the user, and $\mathrm{F} '()$ is the current feature map. 
\clb{Therefore, the position $p_{i}$ is updated to the point within the square region with side length $r_{m}$ around $p_{i}$ in the current feature map that is most similar to the initial point $p_{i}^{0}$.}
After getting the new control points $p_{i}$, we calculate the feature disappearance using Equation~\ref{eq:feature_disappearance}.
\begin{equation}
\mathrm{D} = || {I}_{i}(p_i) - {I}_{0}(p_i^0) ||_2^2
\label{eq:feature_disappearance}
\end{equation}
where ${I}_{0}()$ represents the initial image, ${I}_{i}()$ represents current image. We terminate the iteration when the $\mathrm{D}$ exceeds the threshold.

\section{Theoretical Foundation}
\label{sec:theoreticalFoudation}
\begin{figure}[htb]
  \centering 
  \includegraphics[width=0.9\columnwidth]{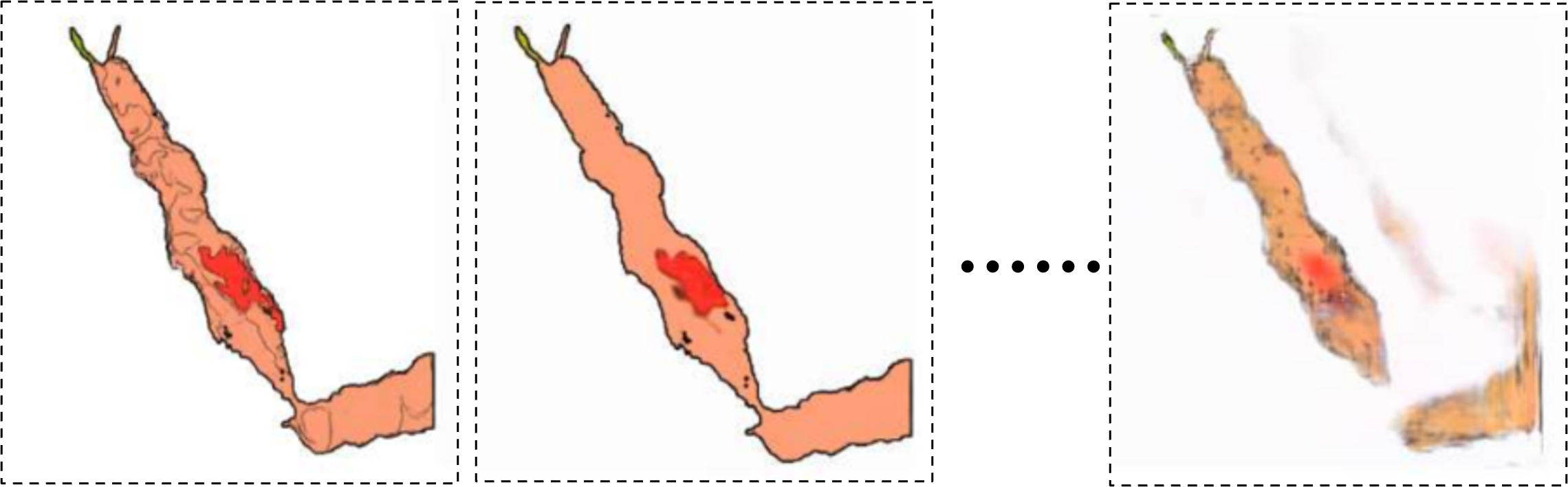}
  \caption{Continuous changes to the latent vectors of \sysname{} generator can lead to the collapse of the generated images.
  }
  \label{fig:image_collapse}
\end{figure}

Our work references the image editing method employed by DragGAN, but directly applying this approach to our generator encountered issues. 
DragGAN generates images by manipulating the generator's latent space vector $W$. The generator used in DragGAN is derived from StyleGAN2, which not only utilizes noise as input but also introduces noise into the latent space for diversity. 
Applying DragGAN's latent space editing technique to our generator resulting in generated images collapsing after only a few iterative modifications to the latent vectors, as demonstrated in Figure~\ref{fig:image_collapse}. 
\clb{In this section, we discuss and provide evidence for the differences in the latent vector distributions and latent vector editing between two models on the same dataset.}

\subsection{Differences in Latent Vector Distributions}
\begin{figure}[htb]
  \centering 
  \includegraphics[width=0.9\columnwidth]{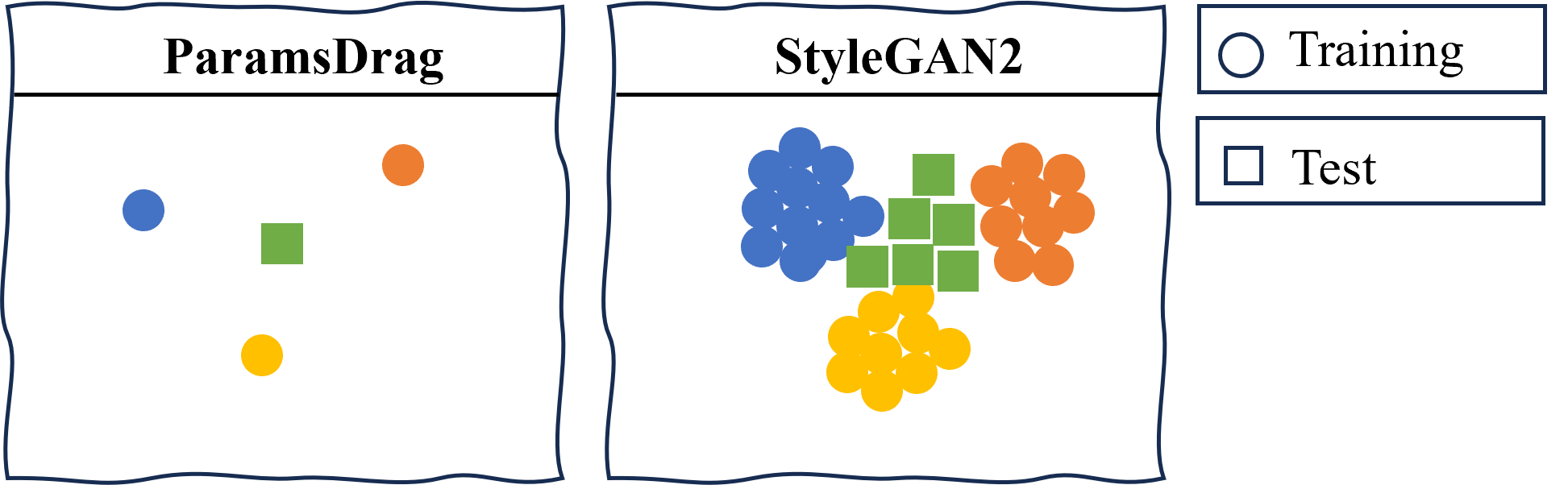}
  \caption{\clb{Schematic representation of the distribution of latent vectors for the same amount of training and test data under two different models.} 
  }
  \label{fig:vec_distribution_with_test}
\end{figure}

\clb{\sysname{} exhibits a distinct distribution of latent vectors compared to StyleGAN2, necessitating different approaches for latent space manipulation. \sysname{} employs explicit input parameters to generate visualization images, establishing a one-to-one correspondence between input and output. Consequently, each individual training data point is associated with a unique vector in the latent space. In contrast, StyleGAN2 employs noise augmentation within the latent vectors, resulting in multiple latent vectors corresponding to a single input parameter, thereby complicating the mapping from input to latent representation. 
Figure~\ref{fig:vec_distribution_with_test} provides a schematic representation of the latent vector distributions for both models, where each dot within the figure represents a sample from the training data. Dots of the same color correspond to the same data sample, with a total of three samples depicted. 
It is evident from the figure that each of the three samples from \sysname{} correlates to a unique latent vector within the latent space. Conversely, StyleGAN2's approach results in each of the three samples correlating to clusters of latent vectors within the latent space.
Moreover, methodologies such as InSituNet, GNN-Surrogate, and VDL-Surrogate similarly adopt explicit labeling strategies to generate images, aligning their latent space vector distributions closely with that of our model.
}

\clb{The latent vector distributions corresponding to the test sets also differs under the two types of models, as shown in Figure~\ref{fig:vec_distribution_with_test}. 
In \sysname{}, each test data point corresponds uniquely to a single latent vector. Conversely, in StyleGAN2, a single test data point is associated with multiple latent vectors. 
\sysname{} demonstrates a discrete distribution of latent vectors in the latent space for both training and test data. In contrast, due to the influence of noise, the latent vectors for the training and test data in StyleGAN2 are clustered and these vectors have similar values. 
The blank areas in the figure also contain numerous latent vectors, which correspond to invalid images. 
Therefore, as the training dataset increases, the latent space of StyleGAN2 will encompass a large number of valid latent vectors and tend to become continuous, while the latent space of \sysname{} remains relatively discrete.
}

It is important to note that although the latent vector distribution of the test set in our model, as seen in Figure~\ref{fig:vec_distribution_with_test}, is near the training set, this does not mean that all points near the training set are meaningful. 
This is because, despite the latent vectors being close in the figure, their distance in high-dimensional space is remote. Moreover, there are many vectors situated between them, the majority of which are invalid.

\subsection{Distinctions in Editing Latent Vectors}
\begin{figure}[htb]
  \centering 
  \includegraphics[width=0.9\columnwidth]{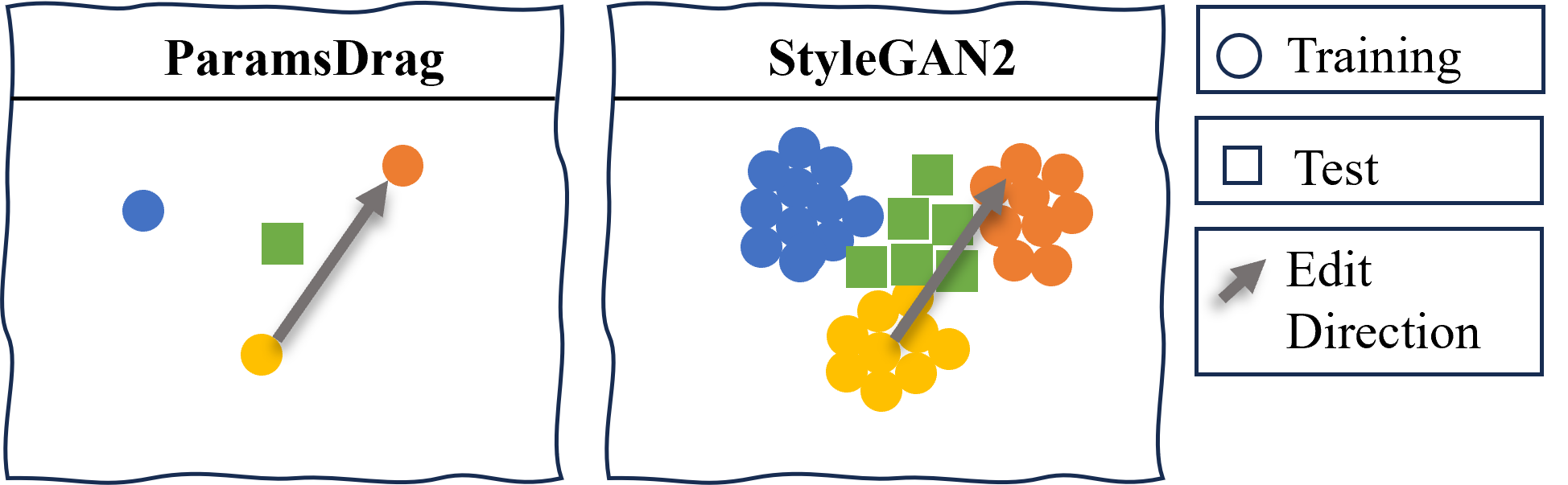}
  \caption{Schematic representation of the differences in continuous changes of latent vectors under two models, with arrows indicating the direction of latent vector editing. \clb{It shows that the continuous changes of latent vectors in discrete space lead to vectors becoming invalid points in blank areas, resulting in image collapse.}
  }
  \label{fig:vec_distribution_with_move}
\end{figure}

The DragGAN approach operates by modifying the latent space vector $W$ to generate new images, utilizing the generator from StyleGAN2. As illustrated in Figure~\ref{fig:vec_distribution_with_test}, the latent vectors of StyleGAN2 exhibit a tendency toward continuity within the latent space, whereas the latent vectors from \sysname{} are discrete. The original DragGAN method proposes a continuous transition from a starting latent vector to a target point; however, such a process under \sysname{}'s discrete latent space may lead to transitions to invalid points. 
\clb{Figure~\ref{fig:vec_distribution_with_move} demonstrates the differences in pass-through of the valid points for continuous editing of latent vectors between the two models. 
It reveals that continuous modifications in a model characterized by a discrete distribution of latent vectors result in transitions to invalid points, producing erroneous images. Conversely, StyleGAN2 features a continuous distribution of latent vectors, enabling seamless and coherent changes within the latent space.}

Overall, generative models under explicit parameter constraints do not continuously edit and move the vector in the latent space. The latent space of \sysname{} exhibits a discrete distribution of valid points. 
Consequently, a straightforward approach is to develop stepwise transitions between these discrete points within the latent space.
Thus, a straightforward strategy is to facilitate \textbf{stepwise transitions between these valid points}, enabling the latent vectors to change in a leapfrog manner. This approach ensures the validity of the generated images.
After training, the \sysname{} can accurately predict visual images based on parameters and provides precise predictions for any input parameters within a reasonable range. This demonstrates that the model can effectively learn the mapping relationship from parameters to visual images, with each input corresponding to a point in the latent space with generalization capabilities. 
Therefore, we can modify the input parameters to induce leapfrog changes in the latent vectors. By implementing a gradient descent search algorithm for input parameters under the constraints of feature supervision and motion supervision, we can effectively solve the challenge of editing latent vectors in a discrete latent space. The theory above guided us in designing the method for editing latent vectors in \sysname{}.

\subsection{Validation on Real Datasets}
\begin{figure}[htb]
  \centering 
  \includegraphics[width=\columnwidth]{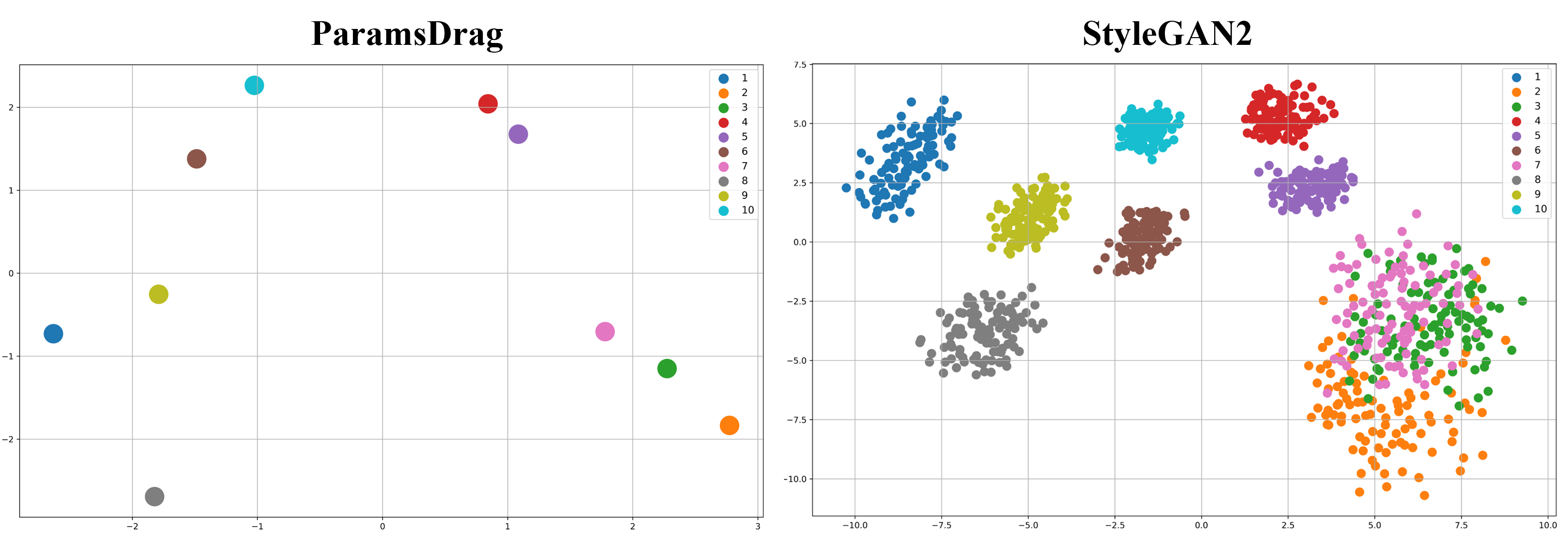}
  \caption{\clb{Distribution maps of latent vectors for 10 samples from the actual dataset under two models.} 
  }
  \label{fig:real_vec}
\end{figure}

To substantiate the validity of the aforementioned theory, we present a visualization of the latent space distribution on a real dataset, as depicted in Figure~\ref{fig:real_vec}. 
\clb{This figure displays the distribution of hidden vectors for 10 data samples under two different models.
The scatter points in the figure represent the latent vectors, and we employ the Multidimensional Scaling (MDS) algorithm to reduce dimensionality and visualize the distribution.}
In Figure~\ref{fig:real_vec}, each point on the left graph represents a latent vector corresponding to an input in \sysname{} model, whereas the right graph displays the latent vectors for the same inputs within StyleGAN2 model. 
An observation from the figure is that \sysname{} model's latent vectors are distributed discretely, in contrast to StyleGAN2 model, where latent vectors are clustered. 
Moreover, as the number of data increases, the latent vectors in StyleGAN2 model tend to form a continuum.

\begin{figure}[htb]
  \centering 
  \includegraphics[width=\columnwidth]{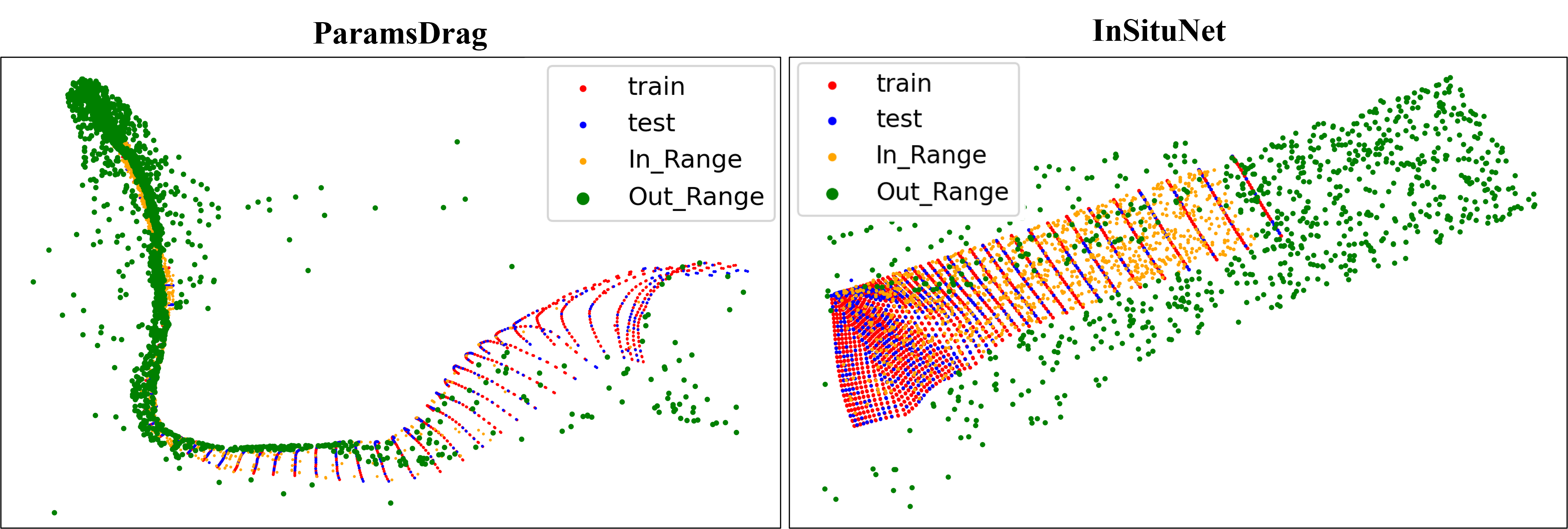}
  \caption{Under parameter constraints, this figure displays the distribution of latent vectors for our model and the InSituNet model. Here, red dots represent the training set, blue denotes the testing set, orange indicates the dataset within the normal parameter range, and green represents the dataset outside the normal parameter range. The dense presence of invalid points (green) among valid ones (orange) suggests easy shifts from validity to invalidity. It also shows that our model has a similar distribution of latent vectors with models like InSituNet, indicating a commonality in generative models for scientific visualization.
  }
  \label{fig:real_params_insitu}
\end{figure}

\clb{We also visualize the distribution of latent vectors corresponding to the same training data, test data, in-range input sampling data, and out-of-range input sampling data for both \sysname{} and InSituNet, as shown in Figure~\ref{fig:real_params_insitu}. The algorithm used for visualizing latent vectors in Figure~\ref{fig:real_params_insitu} employs the MDS algorithm.
In numerical simulations, physical parameters have specific meanings and ranges, and inputs within these correct parameter ranges are defined as valid inputs. Through experimentation, it is observed that images generated from valid inputs are accurate. Sampling points that exceed the range are obtained by sampling far beyond the reasonable parameter limits, and their outputs are typically blurry or chaotic images.
}
It is evident from the figure that the latent vector distributions of both \sysname{} and InSituNet models are discrete. 
The latent vectors corresponding to the test dataset are distributed close to those of the training dataset, and similarly, the latent vectors for the in-range input sampling points are also situated near the training dataset's latent vectors. 
The distribution of the out-of-range input sampling points is more widespread.

\clb{In summary, the valid inputs are clustered around the latent vectors of the training data, yet not all points near these latent vectors are effective. Slight changes in the latent vectors can easily turn them into invalid points. Moreover, if the range of parameters in the training set is much smaller than the actual range of those parameters, this can cause inputs within the normal range of the parameter to become invalid. Therefore, ensuring that the training dataset covers the actual range of the physical parameters is also necessary.}

\section{Experiments}

In this section, we first introduce the experimental setup and then show the effectiveness of our approach.
%\vspace{-1}

\subsection{Experimental Setup}
\subsubsection{Simulation Datasets and Parameters}
In the experiment, we utilized two datasets. The first dataset is the Red Sea simulation, derived from the MIT Ocean general circulation model~\cite{toye2017ensemble}. The second dataset involves the simulation of cosmic large-scale structures sourced from the Gadget-2 simulation program~\cite{springel2005cosmological}. The parameter configurations are described below.

\textbf{Red Sea Simulation:}  
This simulation simulates the circulation dynamics of the Red Sea area. 
The spatial and temporal resolutions are $500 \times 500 \times 50$ and 60 time steps, respectively.
We selected three parameters to conduct the experiments: $Viscosity \in [1.0,4.0]$, $Time \in~[0, 60]$, and $Depth \in [0, 50]$. 
Furthermore, we applied the contour line algorithm to visualize the temperature of each layer to produce images with $256 \times 256$ resolution, where the visualization parameter is fixed throughout the experiment.
\clb{We uniformly sampled 4, 30, and 50 values from viscosity, time, and depth parameters to produce 6000 simulation parameter combinations and render the corresponding images. 
For these data, we randomly selected 5000 as the training dataset and 1000 as the test dataset. During the interaction, the structural similarity parameter $d$ is set to $95\%$, the structural radius parameter $r$ is set to $3$ pixels, the structural displacement parameter $r_m$ is set to $2$ pixels, and the disappearance similarity parameter $D$ is set to $95\%$.}

\textbf{Cosmological  Simulation:} The cosmological simulation employs particles to simulate the evolution of large-scale structures over 13.7 billion years in the universe. 
The parameters and corresponding ranges are $Omega \in [0.05, 1.0]$, representing matter proportion; $OmegaLambda \in~[0, 0.95]$, denoting dark energy proportion; and $Sigma8 \in [0.6, 1.0]$, indicating the power spectrum normalization. We applied a visualization algorithm to convert the particle data into a grid format and output volumetric data with a resolution of $500 \times 500 \times 500$. Then, we used volume rendering to generate global visualizations at a resolution of $512 \times 512$ and surface rendering to generate regional visualizations with the same resolution.
\clb{We uniformly sampled 475 simulation parameter combinations from the parameter space created by the three physical parameters to produce corresponding images.
Subsequently, we randomly selected 380 for the training dataset and 95 for the test dataset.
For the interaction parameters, the structural similarity parameter $d$ is set to $95\%$, the structural radius parameter $r$ is set to $5$ pixels, the structural displacement parameter $r_m$ is set to $8$ pixels, and the disappearance similarity parameter $D$ is set to $95\%$.}

\begin{figure*}[hbt]
  \centering 
  \includegraphics[width=2\columnwidth]{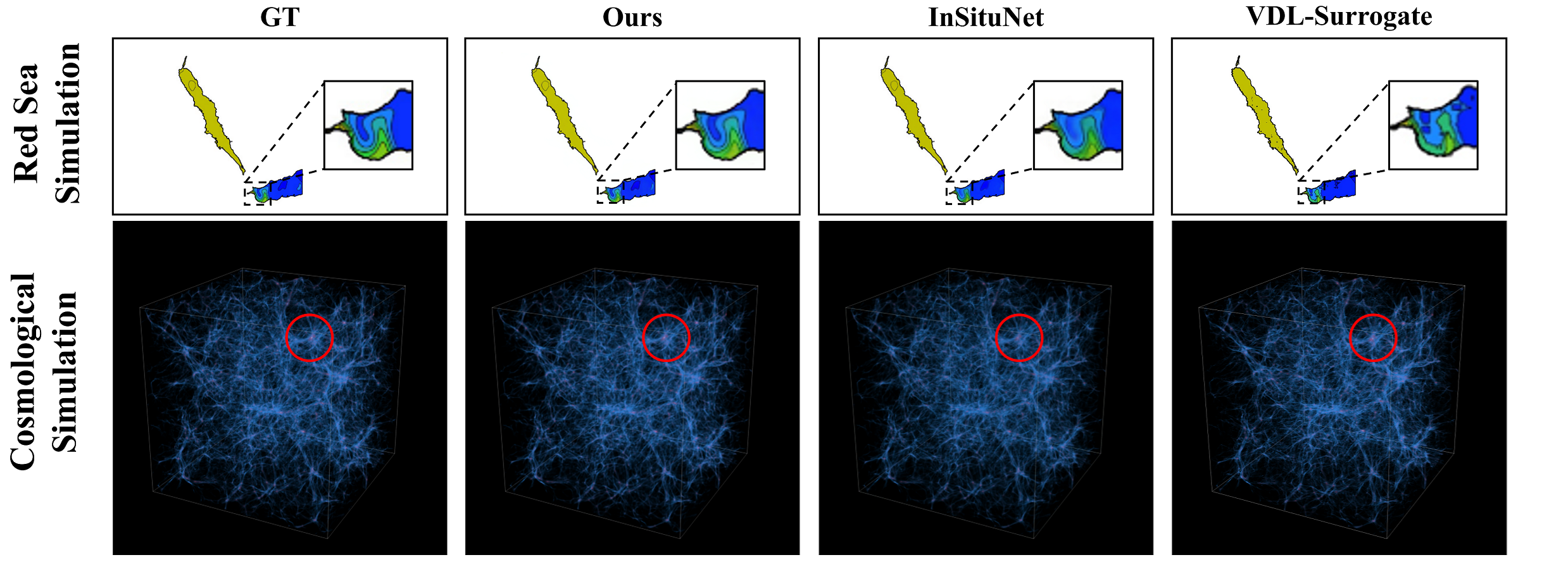}
  \caption{A comparative illustration of our model against InSituNet and VDL-Surrogate models regarding the effectiveness of parameter prediction.  
  }
  \label{fig:case_redsea_pred}
\end{figure*}

\subsubsection{Experimental Environment and Computational Performance}

In the experiments, we utilized a server equipped with two Intel(R) Xeon(R) Gold 6248 processors, two NVIDIA V100 graphics cards, and 128GB of system memory as the environment for model training and prediction. The neural network model implementation is based on the PyTorch Library.
\clb{Our experiments employ the Red Sea dataset and the cosmological simulation dataset. For the Red Sea dataset, our model requires 16 hours for training. For the cosmological simulation dataset, our model training takes 7 hours. The interaction with visualization images is an iterative process, where the speed of each iteration is solely dependent on the image resolution. For images with a resolution of $256 \times 256$ from the Red Sea dataset, an iteration takes $185ms$ on average, whereas for images with a resolution of $512 \times 512$ from the cosmological simulation dataset, an iteration requires $390ms$ on average. The number of iterations needed for one interaction is related to the moving distance of the structure. In our cases, at least 10 iterations are necessary, with a maximum of up to 130 iterations. During the iterative process, the ability to display intermediate images in real-time ensures that users do not perceive any visual delays.}

\begin{figure*}[hbt]
  \centering 
  \includegraphics[width=1.95\columnwidth]{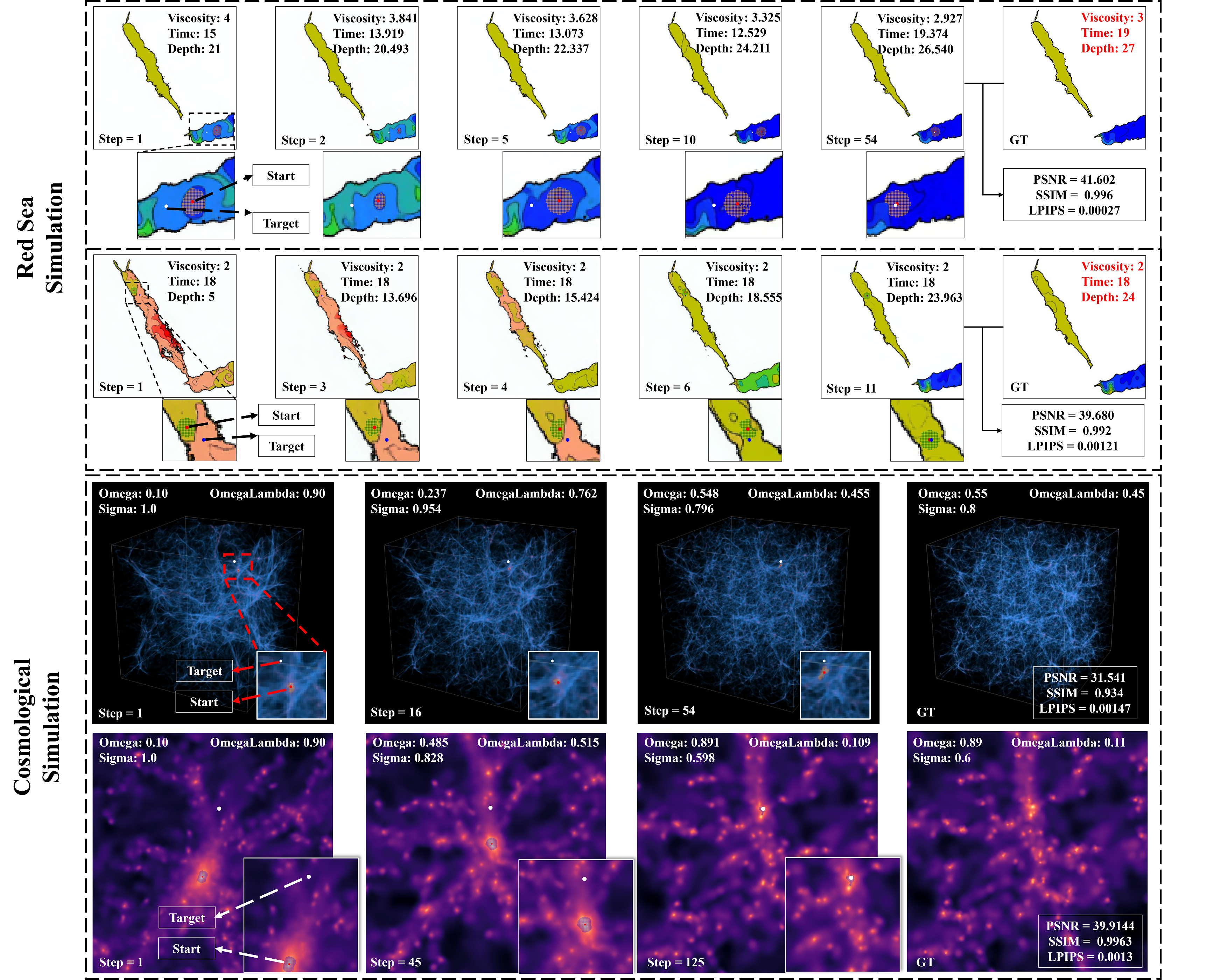}
  \caption{Interactions with the predicted image and the sequence of generated images after the interaction. The parameters used to produce ground truth images are not always identical to the parameters derived from our model. This is because our model always derives floating-point parameters, but simulations could only accept a fixed number of decimal places. Therefore, we round the number to get the ground truth.
  }
  \label{fig:case_interaction}
\end{figure*}

\subsection{Evaluation of Loss Functions}
\clb{Our approach incorporates three distinct loss functions. Table~\ref{tab:tab_loss} presents the mean squared error (MSE) for different combinations of these loss functions across two datasets.
The data indicate that the integration of all three loss functions achieves optimal performance on the Red Sea dataset, whereas the combination of content loss and edge loss is most effective for the cosmological simulation data.
However, the performance difference from using all three loss functions on the cosmological simulation data is minimal, with error margins very close to the best result, making these subtle differences hard to detect in the visual outputs.
Therefore, the combination of the three loss functions is suitable for most applications, and we employed this combination in our experiment.
Regarding the weights $\alpha$, $\beta$, and $\gamma$ for the combined loss functions, experimental results reveal that the error is minimized at $\alpha=1$, $\beta=1$, and $\gamma=0.01$. 
}

\begin{table}[htb]
  \caption{\clb{The comparison of loss functions. The values are the mean squared error between the prediction results and and the raw images from the test dataset.}
  }
  \label{tab:tab_loss}
  \scriptsize%
  \centering%
  \begin{tabu}{%
  	  r%
  	  	*{7}{c}%
  	  	*{2}{r}%
  	}
        \toprule
  	& Loss Functions &  Red Sea & Cosmos \\
  	\midrule
  	& $\mathcal{L}_{content}$ & 0.00153  & 0.00117  \\
  	& $\mathcal{L}_{feature}$ & 0.00088  & 0.00146  \\
        & $\mathcal{L}_{content} + 0.01\mathcal{L}_{edge}$ & 0.00132 & \textbf{0.00114} \\
        & $\mathcal{L}_{feature} + 0.01\mathcal{L}_{edge}$ & 0.00088 & 0.00137 \\
  	& $\mathcal{L}_{content} + \mathcal{L}_{feature} + 0.01\mathcal{L}_{edge}$ & \textbf{0.00081} & 0.00115 \\
  	\bottomrule
  \end{tabu}%
\end{table}

\subsection{Evaluation of Prediction Performance}
\label{sec:evalPredPerformance}
In this section, we qualitatively and quantitatively compared the visualization generation ability of our model with InSituNet and VDL-Surrogate.
Figure~\ref{fig:case_redsea_pred} shows generated images from three models and the ground truth for qualitative comparison. In the Red Sea simulation, the detail generation of our model outperforms the other two approaches. The close-up views reveal that the contour of the isotherms generated by our model is more similar to the ground truth image. Although InSituNet can also correctly generate these details, the image is blurry. The VDL-Surrogate cannot generate the correct isotherms. In the cosmological simulation, the overall visual differences among the three methods are not significant. Only local features present the difference. For instance, the colors of the halos are in red circles. Our method is closer to the ground truth, InSituNet still shows blurred local features, and the VDL-Surrogate result is sharper but loses correct details.

Table~\ref{tab:vis_error} is the quantitative comparison of image generation quality using PSNR (Peak Signal-to-Noise Ratio), SSIM (Structural Similarity Index Measure) and LPIPS (Learned Perceptual Image Patch Similarity).
It shows that the generation ability outperforms alternatives in most of the metrics. Only VDL-Surrogate gets a slightly better SSIM value than ours when generating the cosmological dataset image. 
\clb{
The SSIM is a metric predominantly concerned with the structural integrity of images. Images generated by VDL-Surrogate typically exhibit a reduction in detail and enhanced sharpness of structures, which contributes to their superior performance in SSIM evaluations. In contrast, our method, although less pronounced in the sharpness of structural depiction, more accurately captures details that align closely with the ground truth. As a result, our approach excels in terms of PSNR and LPIPS. 
The core of our method lies in the ability to intuitively manipulate features within visualization images and obtain corresponding simulation parameters, rather than solely focusing on surpassing the performance of existing surrogate models. The results in Table~\ref{tab:vis_error} demonstrate that our model is on par with the state-of-the-art models.}

\begin{table}[htb]
  \caption{Comparison of our model with InSituNet and VDL-Surrogate models across three evaluation metrics on the test dataset.
  }
  \label{tab:vis_error}
  \scriptsize%
  \centering%
  \begin{tabu}{%
  	  r%
  	  	*{7}{c}%
  	  	*{2}{r}%
  	}
  	\toprule
  	& Model     &  PSNR & SSIM  & LPIPS \\
  	\midrule
        \multirow{3}{*}{\makecell{Red Sea \\ Simulation}}
  	& VDL-Surrogate       & 25.251  & 0.954 & 0.0546 \\
        & InSituNet & 38.782  & 0.988 & 0.0041 \\
  	& Ours      & \textbf{41.216}  & \textbf{0.990} & \textbf{0.0035} \\
  	\midrule
        \multirow{3}{*}{\makecell{Cosmological \\ Simulation}}
  	& VDL-Surrogate       & 37.415  & \textbf{0.989} & 0.0040 \\
        & InSituNet & 37.002  & 0.975 & 0.0025 \\
  	& Ours      & \textbf{38.662}  & 0.978 & \textbf{0.0021} \\
  	\bottomrule
  \end{tabu}%
\end{table}

\subsection{Evaluation of Interactive Parameter Exploration}
\label{sec:evalParaInteraction}
Figure~\ref{fig:case_interaction} demonstrates the effects of direct interaction with images. For each image, the corresponding physical parameters are displayed above it, with 'Step' indicating the number of iterations after the interaction, starting from 1 until the stopping condition is met. The 'Start' point represents the user-selected position, surrounded by a mask indicating the selected structural area. The 'Target' point is the intended location, moving the structure from the 'Start' point to the 'Target' point. We also provided corresponding ground truth images for each set of interactions to compare parameters and visualization results.

Figure~\ref{fig:case_interaction} presents two cases from the Red Sea simulation. In the first case (the first row in Figure~\ref{fig:case_interaction}), we moved the area of the lowest temperature in the middle layer of the Gulf of Aden towards the Mandab Strait to observe the changes in three physical parameters and the corresponding visualization images. The generated image sequence reveals that as the cold region expands, the ocean depth and simulation time oscillate while viscosity gradually decreases. Eventually, when the cold region extends to the target location, the ocean depth is lower than its initial state, and the temperature throughout the Gulf of Aden significantly drops under the current parameters. We also show the visualization of the data produced from the simulation by giving the simulation parameter derived by our approach after dragging at the right-most column. The image from our model is nearly identical to the ground truth image.
In the second case (the second row of images in Figure~\ref{fig:case_interaction}), we demonstrate that our approach allows users to fix the viscosity and time parameters and only analyze the relationship between the depth parameter and temperature data. We selected and dragged the cold region at the temperature boundary on the upper side of the Red Sea to examine how the depth changes as these cold regions move downwards. The visualization results show that its depth rapidly decreases as the cold area moves downward. When the selected structure moves to the target position, the ocean depth has reached from the sea surface to the middle layer of the ocean, and at this point, the temperature across the entire Red Sea becomes consistent. The ground truth and the image from our model also show no significant difference. These cases show that our approach allows users to intuitively drag the feature of interest to the desired location with transitions for users to observe the evolution of parameters of interest and images.

In the demonstration of the cosmological simulation data (the third row of images in Figure~\ref{fig:case_interaction}), we dragged the position of the largest halo within the predicted visualization image to observe changes in that halo and overall visualization. The visualization shows that as the position of the halo moves, the global filament structure increases, and the halo gradually dissipates. Ultimately, it is apparent from the visualization that the selected structure did not move to the target position. This is because our model can recognize that the selected structure would not exist at the target location under the constraints of the simulation model. By sending the derived simulation parameters to the computer simulation to get the corresponding data and visualization (the right-most image), an almost identical image is also shown under the parameters.
The second case of the cosmological simulation (the fourth row of images in Figure~\ref{fig:case_interaction}) shows the zoom-in images from the surface rendering visualization. In this case, the selected halo can exist at the target location under the constraints of the simulation model. Therefore, the halo can be moved to the target location. The visualization of the data from the cosmological simulation by the derived simulation parameter from our approach shows an almost identical image. It also verifies that the structure can exist at the target location in the simulation model.

\section{Discussion}
Parameter inversion is a critical component, where typically multiple inputs can map to the same output, yet our approach yields a unique input. This input results from considering the continuity in structural variations. Our model begins with initial input parameters and modifies them such that both parameter and image changes are minimized, achieving a target parameter closest to the original set. This approach aligns with expert needs in image manipulation, aiming for minimal adjustments to produce the desired image. However, allowing for more significant changes during parameter adjustment (e.g., higher learning rates in feature supervision) can produce various target outcomes in the generated sequence of dynamic visualizations. Adjusting the learning rate thus accommodates diverse application requirements.

\clb{Our method can also support higher-resolution image prediction and interaction by adjusting the generator's architecture, specifically by adding synthesis blocks in the Synthesis Subnet to handle increased resolutions. Higher-resolution image processing requires more computational resources and longer training times. 
Additionally, managing larger 3D simulations becomes challenging. This means dealing with larger and more complex data features, which may result in longer training times and less satisfactory prediction outcomes for our method. Therefore, further research in core dataset selection and model generation is essential to improve our method's effectiveness.}

\clb{Our study has identified several limitations. 
First, constructing the training dataset may require multiple iterations if the coverage of physical parameters or sampling frequency is insufficient, leading to poor prediction outcomes and necessitating additional trials to enhance the dataset.
Second, the setup of structure-based patches heavily relies on the application domain, requiring collaboration with domain experts for iterative adjustments. Inappropriate parameter settings can render these patches ineffective at capturing the intended structures. 
Furthermore, in cases with numerous similar local structures, incorrect structure selection and tracking errors can occur during the interactions. 
Lastly, interacting with three-dimensional structures on a two-dimensional plane introduces inaccuracies. The occlusion issues inherent in displaying 3D structures in 2D images complicate interactive selections. While adjusting the viewing angle can partially address this, further research is needed to develop more effective interaction techniques.}

\section{Conclusion}
%In this work, we propose \sysname{}, a surrogate model designed to enhance parameter space exploration via direct interaction with visualizations. We study the distinctions in latent vector distributions and editing methodologies between traditional image generation and scientific data visualization generation. Building on this theoretical groundwork, we develop a surrogate model that uses input simulation parameters to produce relevant scientific data visualizations. Furthermore, we propose a gradient descent-based algorithm for editing latent vectors, enabling stepwise movement on valid points in the discrete space. Additionally, a function retrieves simulation parameters for any image as features are adjusted in the visualization. Through experiments conducted on real-world simulations and comparisons with state-of-the-art solutions, we demonstrate the efficacy of our model.

In this work, we propose \sysname{}, a surrogate model that enhances parameter space exploration through direct interaction with visualizations. We explore differences in latent vector distributions and editing methods between traditional image generation and scientific visualization. We use a gradient descent-based algorithm for editing latent vectors, allowing for precise adjustments in discrete space. Our experiments with real-world simulations show our model's effectiveness compared to state-of-the-art solutions.

%% if specified like this the section will be omitted in review mode
\acknowledgments{%
This work was supported by the National Key Research and Development Program of China (Grant No.2023YFB3002500) and the National Natural Science Foundation of China (No.62202446). The numerical calculations in this study were carried out on the ORISE Supercomputer and the Earth System Numerical Simulation Facility (EarthLab).
}

\bibliographystyle{abbrv-doi-hyperref}

\bibliography{template}

%\appendix % You can use the `hideappendix` class option to skip everything after \appendix

%\section{About Appendices}
%Refer to \cref{sec:appendices_inst} for instructions regarding appendices.

\end{document}